\documentclass[aps,prl,twocolumn,superscriptaddress,footinbib,showpacs,amsfonts,amssymb,amsmath]{revtex4}
\usepackage{graphicx,bm}
\begin{document}
%Title of paper
\title{Glassy Spin Freezing and Gapless Spin Dynamics in a Spatially Anisotropic Triangular Antiferromagnet Ag$_{\textbf{2}}$MnO$_{\textbf{2}}$}

\author{S. Ji}
\affiliation{Department of Physics, University of Virginia, Charlottesville, VA 22904-4714, USA}
\affiliation{NIST Center for Neutron Research, National Institute of Standards and Technology, Gaithersburg, MD 20899, USA}
\author{J.-H. Kim}
\affiliation{Department of Physics, University of Virginia, Charlottesville, VA 22904-4714, USA}
\author{Y. Qiu}
\affiliation{NIST Center for Neutron Research, National Institute of Standards and Technology, Gaithersburg, MD 20899, USA}
\author{M. Matsuda}
\affiliation{Quantum Beam Science Directorate, Japan Atomic Energy Agency (JAEA), Tokai, Ibaraki 319-1195, Japan}
\author{H. Yoshida}
\author{Z. Hiroi}
\affiliation{Institute for Solid State Physics, University of Tokyo, Kashiwa, Chiba 277-8581, Japan}
\author{M. A. Green}
\affiliation{NIST Center for Neutron Research, National Institute of Standards and Technology, Gaithersburg, MD 20899, USA}
\author{T. Ziman}
\affiliation{Institut Laue Langevin, Boite Postale 156, F-38042 Grenoble Cedex 9, France}
\author{S.-H. Lee}
\email{shlee@virginia.edu}
\affiliation{Department of Physics, University of Virginia, Charlottesville, VA 22904-4714, USA}

\date{\today}

\begin{abstract}
Using elastic and inelastic neutron scattering techniques, we show that upon cooling a spatially anisotropic triangular antiferromagnet Ag$_2$MnO$_2$ freezes below $T_f ~\sim 50$ K into short range collinear state. The static spin correlations are extremely two-dimensional, and the spin fluctuations are gapless with two characteristic relaxation rates that behave linearly with temperature.
 \end{abstract}

%\keywords{}
\maketitle

Since a quantum mechanical valence bond state was proposed as the ground state of triangular antiferromagnets in 1970s \cite{anderson73}, triangular lattice systems have been extensively studied, theoretically and experimentally. For the spatially isotropic triangular system, there is now
a consensus that even quantum spins are  magnetically ordered with the 120$^{\circ}$ spiral structure
of the classical state. Recently interests in the field focus on further neighbor interactions, multi-spin interactions, and spatial anisotropy that may lead to more exotic states.\cite{Misguichi, Morita, Merinoetal, Starykh, Nishiyama, HayashiOgata, PardiniSingh} 
The spatially anisotropic  triangular antiferromagnet with stronger exchange coupling $J$ along a chain direction and weaker frustrated zig-zag $J'$ between the chains is interesting because we can also study the effects of the one- to two-dimensional cross-over, but the nature of the ground state is unclear.\cite{Merinoetal, Starykh, Nishiyama, HayashiOgata, PardiniSingh} 

Experimental studies on the spatially anisotropic triangular system have been limited because of scarcity of model systems. The most studied compound is Cs$_2$CuCl$_4$ (s = 1/2) that exhibits a gapless excitation spectrum at low temperatures which was attributed to s = 1/2 spinon excitations.\cite{Coldea1} For $T < 0.17~|\Theta_{CW}|$ ($\Theta_{CW}$ the Curie-Weiss temperature), finite interlayer couplings, $J^{''} = 0.045 J$, drive the system into long range incommensurate spiral order.\cite{Coldea2} For a larger spin, NaMnO$_2$ (s = 2) has recently been shown to have long range collinear order  with a gapped excitation spectrum below $T_N \sim 0.092~|\Theta_{CW}|$.\cite{Giot, Stock} In these systems, the neighboring magnetic triangular planes are separated by a single nonmagnetic layer and weak interplane interactions cause the observed long range orders at low temperatures. 

More recently, Ag$_2$MnO$_2$ was synthesized as a promising candidate for the triangular antiferromagnet where magnetic MnO$_2$ layers are separated by nonmagnetic Ag {\it bi}-layers.\cite{Yoshida08} Bulk susceptibility data showed that the Mn$^{3+}$ ions possess an effective moment of $p_{\textit{eff}}$ = 4.93 $\mu_B$, which is consistent with the high spin s = 2 state of the Mn$^{3+} (t_{2g}^3e_g^1)$ ion. Although $\Theta_{CW} \sim -400$ K, indicating strong antiferromagnetic interactions, no long range order was observed down to 2 K, suggesting strong frustration. Below $T_g$ = 22 K $\sim 0.05~|\Theta_{CW}|$, on the other hand, a field-cooled and zero-field-cooled hysteresis was observed in the bulk susceptibility measurements, which is indicative of a spin freezing.\cite{Yoshida08} 

In this letter, we report our elastic and inelastic neutron scattering measurements on a powder sample of Ag$_2$MnO$_2$. Our principal results are the following. (1) Upon cooling, it undergoes a structural phase transition at 540 K from trigonal to monoclinic due to a ferro-orbital order of the Jahn-Teller active Mn$^{3+}$ ion. This results in spatially anisotropic magnetic interactions in the triangular plane. (2) Despite the large $\Theta_{CW}$, it does not order down to $T_f =$ 48(6) K below which the Mn spins freeze into a collinear spin state with the frozen moment, $\langle M \rangle$ = 2.4(2) $\mu_{\textrm{B}}$/Mn $<< gs \mu_{\textrm{B}}$/Mn.  (3) The frozen spin order is short ranged with anisotropic inplane correlation lengths of $\xi_{b} = 18.9(37)$ \AA~, $\xi_a = 5.9(18)$ \AA~and an out-of-plane correlation length of $\xi_c = 1.6(16)$ \AA, indicating extreme two-dimensionality. (4) The two-dimensional spin fluctuations have a gapless spectrum with two characteristic relaxation rates, an overall relaxation rate, $\Gamma_0$, and a lower limit, $\Gamma_1$, that behave linearly above $T_g$ and $T_f$, respectively. We argue that Ag$_2$MnO$_2$ might be an excellent candidate for a gapless spin liquid phase.

%-----------------------------------
%	Experimental Config. 
%----------------------------------- 
  A 2 g powder sample of Ag$_2$MnO$_2$ was prepared at the ISSP of the University of Tokyo using the solid-state reaction technique with stoichiometric mixture of Ag and MnO$_2$ powder. 
  A series of neutron scattering measurements were performed at the NIST Center for Neutron Research (NCNR).  Time-of-flight neutron scattering measurements were carried out using the disk chopper spectrometer (DCS) with wavelengths of $\lambda =$ 1.8 {\AA}, 2.9 {\AA} and 4.8 {\AA}. 
  Neutron powder diffraction (NPD) measurements were performed on the BT1 powder diffractormeter with a Cu(311) monochromator ($\lambda$ = 1.5403 {\AA}), and Rietvelt refinement was carried out using FULLPROF program \cite{fullprof}. Temperature dependence of the nuclear Bragg peaks was studied at TAS-2 located at the JRR-3 with 14.7 meV incident neutrons and horizontal collimations of guide-80$^{'}$-80$^{'}$-40$^{'}$.

%-----------------------------------------
%	Figure 1.
%-----------------------------------------
\begin{figure}
\includegraphics[width=0.5\textwidth]{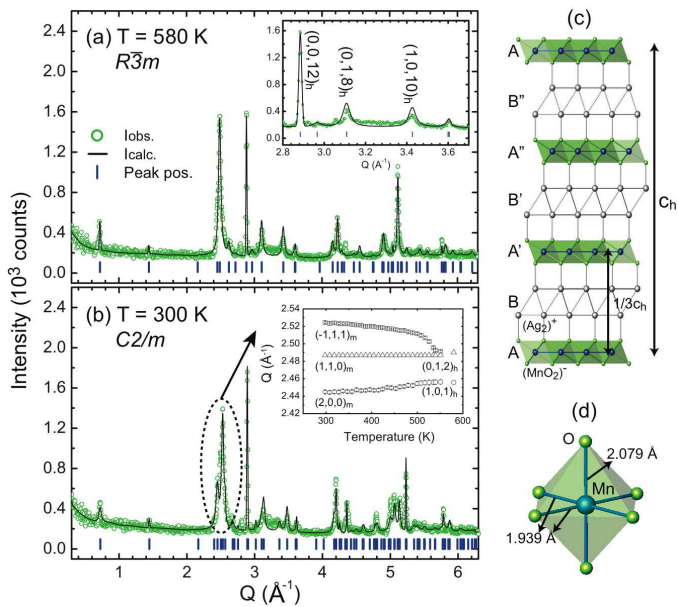}
\centering
\caption{(Color online) Neutron powder diffraction data measured (a) at 580 K and (b) 300 K. Circles are the data and the line represents the calculated intensity based on the lattice parameters listed in Table I. The inset of (a) shows a close-up of a narrow range of wavevector $Q$ while the inset of (b) shows the monoclinic splitting of the two peaks around $Q \sim 2.5$ \AA$^{-1}$ into three peaks below 540 K. (c) shows the stacking of the MnO$_2$ and Ag layers along the $c$-axis in the high temperature trigonal phase. (d) shows the local Jahn-Teller distortion of the MnO$_6$ octahedron.
\label{fig3}}
\end{figure}

As shown in Fig. 1 (a), at 580 K the nuclear Bragg reflection positions tell us that the high temperature crystal structure is trigonal with $R\bar{3}m$ symmetry. The (0,0,L) reflections are instrument resolution limited, however, the (H,K,L) reflections with nonzero H or K are much broader than the Q-resolution. 
The best fit as shown as the line was obtained with the lattice parameters listed in Table I and the stacking correlation length of 217(37) \AA. As shown in Fig. 1 (c), the chemical unit cell of the perfect hexagonal structure consists of three MnO$_2$ layers (A, A$^{'}$ and A$^{''}$) and three Ag bi-layers (B, B$^{'}$ and B$^{''}$) that appear alternately. Neighboring layers of same kind are displaced by (1/3,1/3,1/3). The stacking faults may occur due to weak Ag-O van der Waals and ionic bondings between the Ag bi-layer and the neighboring MnO$_2$ layer (see Fig. 1 (c)). Thus, the stacking order of the layers can be imperfect: instead of the long range stacking of A-B-A$^{'}$-B$^{'}$-A$^{''}$-B$^{''}$ as expected in a perfect crystal, stacking faults such as A-B-A$^{'}$-B$^{''}$-A$^{''}$-B$^{'}$ or A-B$^{''}$-A$^{''}$-B-A$^{'}$-B$^{'}$ may occur. Such stacking faults will not change the $c$-positions of the layers but disorder the arrangements of the $ab$-positions of the atoms along the $c$-axis, and yield the observed broadenings of the $(H \neq 0,K \neq 0,L)$ nuclear Bragg reflections. %Ironically, such stacking faults will further suppress magnetic interlayer coupling between the MnO$_2$ layers.
%-------------------------------------------------
%	Table I
%-------------------------------------------------
\begin{table}
  \caption{The crystal structural parameters of Ag$_2$MnO$_2$ obtained at 580 K and 300 K by refining the data shown in Fig. 1 using the program Fullprof. }
  \begin{ruledtabular}
    \begin{tabular}{llll}
      Atom($W$) & $x$ & $y$ & $z$ \\ \hline 
      \multicolumn{4}{l}{580 K ($R\bar3 m$), $\chi^2 = 3.92 $, R$_{F^2}$ = 12.1} \\ 
      \multicolumn{4}{l}{$a$ = $b$ = 2.96991(18)\AA, $c$ = 26.14007(229)\AA}\\
      Ag (6c) & 0 & 0 & 0.21064(20) \\
      Mn (3a) & 0 & 0 & 0 \\
      O (6c)  & 0 & 0 & 0.29552(26) \\ 
      \multicolumn{4}{l}{Mn-O = 1.9831(4) {\AA}}\\ \hline
      
      \multicolumn{4}{l}{300 K ($C2/m$), $\chi^2 = 5.48 $, R$_{F^2}$ = 15.0} \\
      \multicolumn{4}{l}{$a$ = 5.24722(60)\AA, $b$ =  2.88226(16)\AA, $c$ = 8.89877(99)\AA}\\
      \multicolumn{4}{l}{$\beta$ = 102.39862(1300)$^{\circ}$}\\
      Ag (4i) & 0.21183(206) & 0 & 0.62875(73) \\
      Mn (2a) & 0 & 0 & 0 \\
      O (4i)  & 0.30348(229) & 0 & 0.88331(91) \\ 
      \multicolumn{4}{l}{Mn-O (apical) = 2.079(6) {\AA}, Mn-O (plane) = 1.939(2) {\AA}} 
    \end{tabular} 
  \end{ruledtabular}
\end{table}

Upon cooling from 580 K, the two Bragg reflections over 2.4 \AA$^{-1} < Q < 2.6$ \AA$^{-1}$ split into five peaks at $\sim$ 540 K, indicating lowering of the crystal symmetry (see the inset of Fig. 1 (b)). %The neutron powder diffraction data measured at 300K is shown in Fig. 1 (b). 
The best refinement of the NPD data taken at 300 K (Fig. 1 (b)) was obtained with a monoclinic $C2/m$ crystal structure with the lattice parameters listed in Table I. The lattice distortion is due to Jahn-Teller distortion of the MnO$_6$ octahedron that involves elongation of a local axis that is close to the $a$-axis, as shown in Fig. 1 (d) and Fig. 3 (b). As a result, the $e_g$ electrons of Mn$^{3+}$ ions occupy $d_{3r^2-z^2}$ orbital, resulting in a ferro-orbital order (see Fig. 3 (b)). %Its implication for the magnetic interactions will be discussed later.

%-------------------------------------------------
%	Figure 2.
%-------------------------------------------------
\begin{figure}
\includegraphics[width=0.5\textwidth]{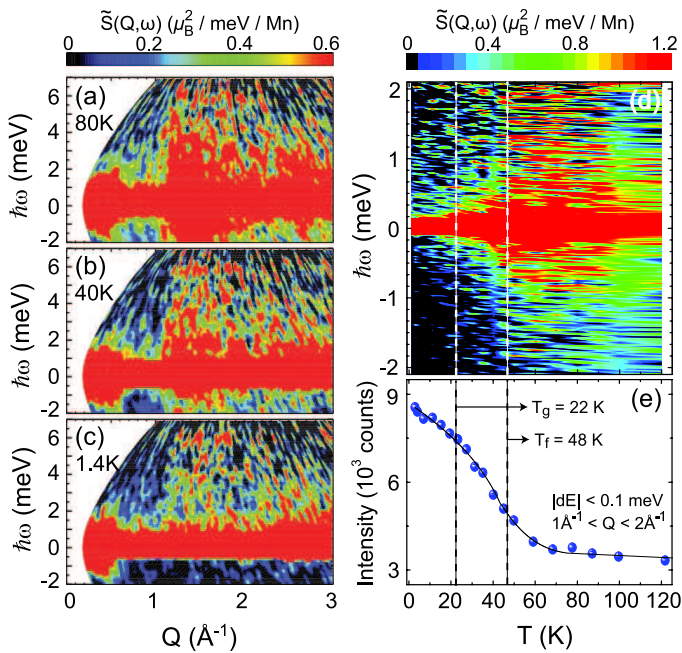}
\centering
\caption{(Color online) (a)-(c) Color contour maps of neutron scattering intensity as a function of wave vector, $Q$, and energy, $\hbar\omega$, transfers, measured with $\lambda = 2.4$ \AA~at (a) 80 K, (b) 40 K, and (c) 1.4 K. (d) $\hbar\omega$-dependence of low energy spin fluctuations measured with $\lambda = 4.9$ \AA~at various temperatures spanning the phase transition. (e) $T$-dependence of the elastic magnetic intensity obtained by the neutron intensity over 1 \AA$^{-1} < Q < 2$ \AA$^{-1}$ and $|dE| <0.1$ meV. $T_f$ was determined from Fig. 4 (d).
\label{fig1}}
\end{figure}

Fig. 2 (a)-(c) show the neutron scattering intensity obtained at DCS as a function of the momentum($Q$) and energy ($\hbar\omega$) transfers, measured at $T$ = 80 K, 40 K and 1.4 K.
  At 80 K $\gg$ $T_\textrm{g}$, $S(\bm{Q},\omega)$ exhibits a broad continuum over $\hbar\omega$.  
  The $Q$-dependence of the low energy continuum is asymmetric with a sharp increase at Q $\simeq$ 1.25 \AA$^{-1}$ and a broad tail at higher $Q$.  This indicates that the spin fluctuations are low-dimensional in nature. 
  %As $T$ decreases and approaches $T$ $\simeq$ $T_\textrm{g}$, the anymmetric lineshape remains but spectral weight of $S(\bm{Q},\omega)$ shifts to lower energies.
  %At 1.5 K $< T_g$, the low energy spin fluctuations become weaker.
  Fig. 2 (d) shows $T$-dependence of the low energy fluctuations $S(\omega) = \int_{1\textrm{\AA}^{-1}}^{2\textrm{\AA}^{-1}} S(Q,\omega) dQ$ obtained from the $\lambda$ = 4.8 {\AA} data with an instrumental energy resolution of $\Delta E =$ 0.11 meV. 
  Upon cooling from 120 K, the low energy spin fluctuations increase and become strongest at 60 K below which they weaken.  Fig 2 (e) shows that the nominally static spin correlations with life time longer than $\Delta \tau_\textrm{min} \simeq \frac{\hbar}{\Delta E}\simeq$ 3.29 ps develop below $T_\textrm{f}$ that is higher than $T_g = 22$ K determined by the bulk susceptibility measurements with $\Delta E$ = 0.
  Detecting different transition temperatures with different energy resolutions is common for magnetic systems such as spin glasses where spins freeze into a short-range ordered state. \cite{Mydosh, shl96_2, msr}
  
%-----------------------------------------
%	Figure 3.
%-----------------------------------------
\begin{figure}
\includegraphics[width=0.5\textwidth]{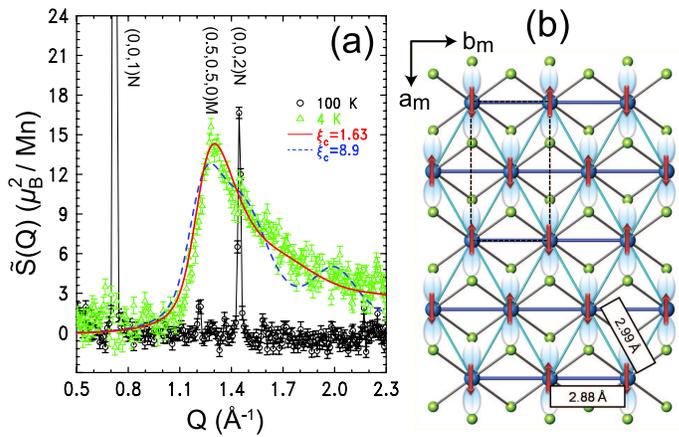}
\centering
\caption{(Color online) (a) Elastic neutron scattering intensity, $\tilde{I} (Q)$, at 4 K and 100 K. For 4 K $<$ T$_f$, nonmagnetic background measured at 100 K was subtracted to get magnetic contributions only. The reflection indices are in the monoclinic notations. Lines are described in the text.  (b) $ab-$projection of the MnO$_2$ triangular layer with a magnetic structure that is consistent with the characteristic wave vector, ${\bf q}_m = (0.5,0.5,0)_{mono}$, of $\tilde{I} (Q)$. Bigger and smaller spheres are Mn and O ions, respectively. Ovals represent the $d_{3r^2-z^2}$ orbitals with the local $z$-axis being close to the $a$-axis. In addition, each Mn$^{3+}$ ion has three $t_{2g}$ orbitals occupied that point between the oxygens. Thick (thin) blue lines are short (long) bonds between neighboring Mn ions. Dotted lines represent the chemical unit cell of the monoclinic phase.
\label{fig2}}
\end{figure}

  In order to understand the nature of static short-range magnetic order below $T_f$, we have plotted $Q$-dependence of the elastic scattering intensity obtained by integrating the $\lambda =$ 4.8 \AA~data over an energy window of $\vert \hbar\omega \vert <$ 0.2 meV.
  Fig. 3 (a) shows the resulting $S(Q)$ at $T$ = 4 K and 100 K. The 100 K $(> T_f)$ data were taken as nonmagnetic background and were subtracted from the other $T$ data. The resulting magnetic $S(Q)$ exhibits a broad peak without any magnetic Bragg peaks, indicating that the static spin correlations are short ranged. The broad peak is peaked at $Q$ = 1.251 {\AA} that corresponds to a characteristic wave vector of ${\bf q}_m =$ (1/2,1/2,0). Furthermore, the elastic $S(Q)$ is asymmetric as the low energy excitations are. These mean that the short range magnetic ordered structure is collinear and low-dimensional, as shown in Fig. 3(b). For a quantitative analysis, we have fitted $S(Q)$ to 
  the elastic neutron cross section described by the product of the independent lattice-Lorentzian functions \cite{IZ_SHL},
\begin{eqnarray}
  \frac{d\sigma_{el}}{d\Omega} (\bm{Q}) \propto |S^{\bot} (\bm{Q})|^2 \prod_{\alpha}
  \frac{\sinh \xi^{-1}_{\alpha}}{\cosh \xi^{-1}_{\alpha} - \cos( (\bm{q}_m-\bm{Q})\cdot \hat{\bm{r}}_{\alpha})}.
  \label{eqn1}
\end{eqnarray}
Here $S^{\bot}(\bm{Q})$, the unit cell magnetic structure factor normal to the scattering vector, can be written as $S^{\bot}(\bm{Q}) =  F(Q)^2 \sum_\nu \bm{M}_\nu^{\bot} e^{-i\bm{Q}\cdot \bm{r}_\nu}$ where $\bm{M}_\nu$ and $\bm{r}_\nu$ are the staggered magnetic moment and position of Mn$^{3+}$ ion at site $\nu$, respectively, and $F(Q)$ is the Mn$^{3+}$ magnetic form factor.  $\xi_{\alpha}$ and $\hat{\bm{r}}_{\alpha}$ are the spin correlation length and the unit lattice vector along the $\alpha$-axis, respectively. The best fit shown as the red solid line in Fig. 3 (a) was obtained with $\xi_{b} = 18.9(37)$ \AA~along the chain, $\xi_a = 5.9(18)$ \AA~perpendicular to the chain in the triangular plane, and a negligible out-of-plane correlation length of $\xi_c = 1.6(16)$ \AA. For comparison, we also show the calculated $S(Q)$ obtained with the same $\xi_b$ and $\xi_b$ but $\xi_c$ to be the interlayer distance between the neighboring triangular layers, $8.9$ \AA~(see the blue dashed line), which does not reproduce the data. These clearly indicate that the magnetic interactions in Ag$_2$MnO$_2$ are extremely two-dimensional, as expected by the large distance between the MnO$_2$ layers that are separated by the non-magnetic Ag {\it bi}-layers. The anisotropic inplane correlation lengths can be understood when the orbital state of Mn$^{3+}$ ($t_{2g}^3 e_g^1$) ions are considered.\cite{Goodenough60, Samuelsen70} As shown in Fig. 3 (b), in the ferro-orbital state the $e_g$ electrons do not induce any obvious superexchange paths between Mn-Mn ions. On the other hand, due to the edge-sharing network of the MnO$_6$ octahedra neighboring $t_{2g}^3$ electrons directly overlap, inducing strong nearest neighbor (NN) interactions that is sensitive to the distance, $d$, between the Mn ions. The lattice distortion shown in Fig. 3 (b) leads to stronger intrachain exchange coupling $J$ with $d \simeq 2.88$ \AA~ and four weaker zig-zag coupling $J^{'}$ with $d \simeq 2.99$ \AA~as shown by dark and light blue lines, respectively. 

 %Why is the ordered spin structure is collinear instead of the 120$^o$ spin configuration that is the ground state for a uniform triangular lattice? COnsidering the nearest neighbor interactions only, the mean-field energy per a triangle with the $J_1$ and $J_1^{'}$ becomes $E_c = -J_1 <S>^2$ and $E_{120} = -\frac{1}{2} (J_1+2 J_1^{'}) <S>^2$ for the collinear and the 120 degree structure, respectively. Thus, the observed short range collinear spin structure indicates that $J_1$ is stronger than $J_1^{'}$ in Ag$_2$MnO$_2$. 

%-----------------------------------------
%	Figure 4.
%-----------------------------------------
\begin{figure}
\includegraphics[width=0.5\textwidth]{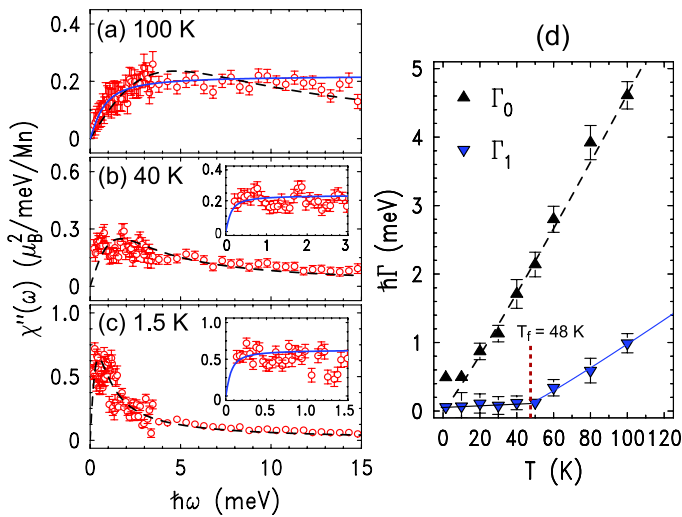}
\centering
\caption{(Color online) (a)-(c) Energy dependence of the imaginary part of the dynamic susceptibility obtained by integrating and converting the inelastic neutron scattering intensity $I(Q,\omega)$ shown in Fig. 2 over 1 \AA$^{-1} < Q <$ 2 \AA$^{-1}$ at (a) 100 K, (b) 40 K, and (c) 1.5 K. (d) Relaxation rate, $\Gamma$, as a function of temperature. Lines are described in the text. \label{fig2}}
\end{figure}

Let us now turn to the nature of the dynamical spin correlations. We obtained energy dependence of the scattering intensity, $I(\omega)$, by integrating all $I(Q,\omega)$ data taken with $\lambda =$ 1.8 {\AA}, 2.8 {\AA} and 4.9 {\AA} over 1 \AA$^{-1} < Q <$ 2 \AA$^{-1}$. Then, using the detailed balance relation $\chi^{''} (\omega) = \frac{\pi}{3} (1-\exp^{-\hbar\omega/k_B/T}) I(\omega)$ where $k_B$ is the Boltzmann constant and the imaginary part of the dynamic susceptibility, $\chi{''}$, was extracted. As shown in Fig. 4 (a), at $T_f <$ 100 K $\ll \vert \Theta_{CW}\vert$,  $\chi^{''} (\omega)$ can be well fitted to a lorenzian, $\chi^{''} (\omega) \propto \Gamma_0\omega/(\Gamma_0^2+\omega^2)$. When temperature decreases, however, the spectral weight shifts down to lower energies and the lorentzian cannot reproduce the low energy region while it fits the higher energy region. Below $T_f$, the low energy region can be fit to $\chi^{''} (\omega) \propto {\rm tan^{-1}} (\omega/\Gamma_1)$ that represents spin relaxations with a distribution of the relaxation rates with the lower limit being $\Gamma_1$.\cite{Mydosh}  The optimal relaxation rates are plotted in Fig. 4 (d). For $T > T_g$, the overall relaxation rate $\Gamma_0 = C_0 (k_B T)^{\alpha_0}$ with $C_0 = 0.5(1)$ and $\alpha_0 = 1.08(16)$. For $T > T_f$ K, the lower limit $\Gamma_1 =  C_1 (k_BT)^{\alpha_1}$ with $C_1 = 0.18(6)$ and $\alpha_1 = 1.07(16)$. 
For $T < T_f$, $\Gamma_1 = 0.11(2)$ meV that is almost zero, independent of temperature. This contrasts with the behavior of $\Gamma$ found in a well-known quasi-two-dimensional system SrCr$_{9p}$Ga$_{12-9p}$O$_{19}$ where the magnetic Cr$^{3+}$ ions form a [111] slab of a three-dimensional network of corner-sharing tetrahedra \cite{shl96_1}: in SCGO, upon cooling above $T_f$, $\Gamma$ decreases linearly to zero at $T_f$, but it increases back upon further cooling, which was attributed to the absence of local low-energy excitations in the frozen state.\cite{shl96_2}

The gapless short range collinear spin order observed in Ag$_2$MnO$_2$ is quite different from the ground states observed in other related materials, Cs$_2$CuCl$_4$ with  gapless long range incommensurate spiral order, and NaMnO$_2$ with gapped long range collinear order. What determines the particular ground state in a spatially anisotropic triangular system? Spin wave analysis for varying the spatial anisotropy $\alpha= {J'\over J}$, \cite{Merinoetal} predicts two regions where the classical ordered state is unstable: for small $\alpha$ where  one-dimensional fluctuations become important and near $\alpha= 1/2$ where classically there is a transition between collinear and spiral phases. The nature of the ordered states developed by quantum fluctuations in these regions is not clear. For s = 1/2,  predictions vary: a collinear magnetic state \cite{Starykh} or a valence bond state, either gapped \cite{Nishiyama} or ungapped \cite{HayashiOgata}  separated by a quantum critical point from the spiral phase. For s = 2, there should be a quantum critical point at  smaller $\alpha$ from the Haldane gap state, as in s = 1 \cite{PardiniSingh}, to either a collinear or a spiral state. While the microscopic origin is not yet certain, the gapless excitations and the glassiness observed in Ag$_2$MnO$_2$ suggest that the collinearity  is an intrinsic property of the anisotropic triangular antiferromagnet. Multiple (three- or four-) spin exchange on the triangle, as dominates the magnetism of absorbed He$^3$ \cite{Roger}, might stabilize  the gapless collinear structure over a wider range of  parameter values \cite{Nishiyama}, and produce a gapless spin liquid state, as proposed for  spin 1/2 \cite{Misguichetal}.

%----------------------------------------
%	Conclusion and Acknowledgements
%----------------------------------------

\begin{acknowledgments}
 This work was supported by NSF under Agreement No. DMR-0903977 and No. DMR-0454672. SHL thank the WPI-Advanced Institute for Materials Research at Tohoku University for their hospitality during his stay when this paper was partially written.

\end{acknowledgments}

% Create the reference section using BibTeX:
\bibliography{ag2mno2_rev8.bbl}

\end{document}